\documentclass[11pt]{article}
\usepackage{epsf}
\usepackage{amsmath}
\usepackage{graphics}
\usepackage{cite}

\renewcommand{\thefootnote}{\#\arabic{footnote}}
\begin{document}

\newcommand{\gtrsim}{ \mathop{}_{\textstyle \sim}^{\textstyle >} }
\newcommand{\lesssim}{ \mathop{}_{\textstyle \sim}^{\textstyle <} }

\newcommand{\rem}[1]{{\bf #1}}

\renewcommand{\thefootnote}{\fnsymbol{footnote}}
\setcounter{footnote}{0}
\begin{titlepage}

\def\thefootnote{\fnsymbol{footnote}}

\hfill February 2018\\
\vskip .5in
\bigskip
\bigskip

\begin{center}
{\Large \bf Maximum PBH Mass and Primordiality}

\vskip .45in

{\bf P.H. Frampton\footnote{email: paul.h.frampton@gmail.com
~~homepage: www.paulframpton.org}}

{Dipartimento di Matematica e Fisica "Ennio De Giorgi"\\
 Universit\`a del Salento and INFN Lecce, 
 Via Arnesano 73100 Lecce,Italy.}
 
\begin{center}
and
\end{center}

{9a South Parade, Oxford OX2 7JL, UK}

\end{center}

\vskip .4in
\begin{abstract}
\noindent
In order to avoid unacceptable $\mu$-distortions inconsistent with 
observational data on the Cosmic Microwave Background,
Primordial Black Holes (PBHs) must be less massive than
$10^{12} M_{\odot}$, quite closely above the highest
black hole mass yet observed. This comparableness leads us 
to posit that all supermassive black holes originate as PBHs.
\end{abstract}

\end{titlepage}

\renewcommand{\thepage}{\arabic{page}}
\setcounter{page}{1}
\renewcommand{\thefootnote}{\#\arabic{footnote}}

\newpage

\noindent
{\it Introduction.}
Surely the most impressive prediction of general relativity (GR) theory is the existence of
black holes in spacetime. The Schwarzschild solution of GR was discovered in
1916 \cite{Schwarzschild} describing a static spherically-symmetric black hole.  
Because of the non-linearity of Einstein's equations, it took until 1963 before Kerr\cite{Kerr}
discovered a solution of GR which described a rotating axially-symmetric
black hole. There exist more general such solutions with electric charge but we shall
assume that all the astrophysical black holes are electrically neutral so the
Kerr solution is the most general classical solution needed.

\bigskip

\noindent
One special property of Kerr black holes is commonly called the no-hair theorem
\cite{Carter} which states that they are completely characterised by mass and spin.
This implies that there is no way of telling how a black hole was formed, whether
primordially or by gravitational collapse of a pre-existing object. Nevertheless, the black holes formed by gravitational
collapse can be formed only after the first stars are formed at about $t \sim 10^8$ yr
(red shift $Z\sim27$) while primordial black holes are formed before $t\sim 10^7$ seconds
(red shift $Z\sim 2\times 10^6$).

\bigskip

\noindent
Stars have masses below $1,000 M_{\odot}$ and therefore can collapse into
black holes only lighter than this. Of course, heavier black holes can be produced by
accretion and mergers of lighter black holes, although it seems extremely unlikely that a
supermassive black hole like the one near the centre of gravity of the Milky Way
called SagA* which has mass $M\sim 4\times 10^6M_{\odot}$ could have been so formed.
Surely SagA* was rather seeded much earlier in the expansion era as we shall advocate in this Letter?

\bigskip

\noindent
Let us begin by studying one thing which is certain, that the CMB spectrum is
extraordinarily close to the black-body Planckian formula for frequency $\nu$
and temperature $T$:

\bigskip

\begin{equation}
F(\nu,T) = \left( \frac{2 h \nu^3}{c^2} \right) \left[ \exp (h\nu/kT) - 1 \right]^{-1}
\label{Planckian}
\end{equation}

\noindent
applicable when the electron-photon plasma is fully in thermal
equlibrium, as happens when the Compton and Double Compton scatterings
are faster than the cosmological expansion. Eq.(\ref{Planckian}) famously
agrees with the measured CMB better than any terrestrially-measured black-body spectrum.

\bigskip

\noindent
As a comparison to Eq.(\ref{Planckian}) absence of perfect thermal equilibrium
in the electron-photon plasma can lead to a distorted 
CMB spectrum \cite{PZ} which can be parametrised {\it e.g.} by

\begin{equation}
F(\nu,T) = \left( \frac{2 h \nu^3}{c^2} \right) \left[ \exp ([h\nu/kT] + \mu) - 1 \right]^{-1}
\label{muDistortion}
\end{equation}

\noindent
in which the chemical potential $\mu$ is strongly constrained, by the aforementioned
accuracy of agreement with Eq.(\ref{Planckian}), 
to $\mu \leq 10^{-4}$ while planned experiments aim for $\mu \leq 10^{-9}$;
a careful analysis is provided in \cite{NCS}.

\bigskip

\noindent
{\it Maximum PBH Mass.}
The mass of a PBH is tied to the horizon size at the time of PBH formation. At 
cosmological time $t$ the PBH mass is given within an order of magnitude by

\begin{equation}
M_{PBH} \simeq 10^5 M_{\odot}  \left( \frac{t}{1 ~ {\rm second}} \right)
\label{Mpbh}
\end{equation}

\noindent
and therefore the maximum $M_{PBH}$ depends on the maximum time
of PBH formation.

\bigskip

\noindent
In order that the CMB spectrum be given by Eq.(\ref{Planckian}) rather than
by Eq.(\ref{muDistortion}) with an unacceptably large $\mu$, the PBH
formation must take place while the electron-photon plasma remains in excellent
thermal equilibrium. There is a cosmic time, called the thermalisation time $t_{th}$,
after which the Compton and Double Compton scattering cannot keep up
with the Hubble expansion so that thermal equilibrium becomes 
unacceptably inexact.

\bigskip

\noindent
Assuming, as is justified {\it a posteriori}, that the thermalisation time
occurs during the radiation-dominated era, its value is calculable with
sufficient precision to cite a $t_{th}$ accurate to an order of magnitude
\cite{NCS} so ignoring factors of order one we shall adopt here the value

\begin{equation}
t_{th} = 10^7 ~ {\rm seconds}
\label{tThermalisation}
\end{equation}

\noindent
which implies an upper limit on PBH mass:

\begin{equation}
M_{PBH} \leq 10^{12} M_{\odot}
\label{PBHlimit}
\end{equation}

\bigskip

\noindent
{\it Observed Supermassive Black Holes}
As already mentioned, the supermassive black hole (SMBH) at the
centre of the Milky Way (MW), SagA*, has an unusually light mass for such
a SMBH at a galactic centre:

\begin{equation}
M^{(MW)}_{SMBH} = M_{SagA^*} \sim 4 \times 10^6 M_{\odot}
\label{Msag}
\end{equation}

\noindent 
The most massive known SMBH at a galactic centre is in galaxy NGC1277
with

\begin{equation}
M^{(NGC1277)}_{SMBH} \sim 1.7 \times 10^{10} M_{\odot}
\label{NGC1277}
\end{equation}

\bigskip

\noindent
One cannot help noticing that the maximum PBH mass expressed in
Eq.(\ref{PBHlimit}) is not much above the maximum so far observed
SMBH mass given for NGC1277 in Eq.(\ref{NGC1277}). We take
this coincidence seriously and not as accidental.

\bigskip

\noindent
{\it Remarks on Primordiality.}
The black hole resulting from the gravitational collapse of a star
cannot be more massive than the original star. If Population III
stars once existed their masses are taken to be in a range
up to a maximum of $\sim 1000M_{\odot}$. These would be the
most massive stars which have ever existed so a maximum
value for gravity collapse black holes, at the time
of their formation and ignoring all subsequent mergers and
accretion is

\begin{equation}
\left\{M^{(collapse)}_{BH} \right\}_{initial} ~~~  \leq ~~~10^3 M_{\odot}
\label{collapse}
\end{equation}

\noindent
Based on Eq.(\ref{collapse}), the subsequent increase of mass of
a gravity-collapsed black hole from $Z=27$ to $Z=0$
to reach the mass of sagA* in
Eq.(\ref{Msag}) would be by a factor of thousands which is impossible
 to underwrite by accretion and merger processes. In the case of 
NGC1277 the needed increase of mass to reach
Eq.(\ref{NGC1277}) by accretion and merging would be by a factor
of {\it ten million} which, without needing a calculation, 
is impossible. Our conclusion is that :\\
\underline{{\bf supermassive black holes are primordial or at least seeded by primordial black holes.}}

\bigskip

\noindent
Let us conclude with some remarks about stellar mass
black holes. Recently, the most exciting development was
the discovery of gravitational waves from merger in a binary
of stellar mass black holes at 410pc ($Z\sim 0.09$) from Earth with 
approximate masses:

\begin{equation}
36 M_{\odot} +29M_{\odot}  \rightarrow  62 M_{\odot} + \left( 3M_{\odot} ~
{\rm in~gravitational~waves} \right)
\label{LIGO}
\end{equation}

\noindent
The energy emitted in gravitational waves is $3 M_{\odot} \equiv 5 \times 10^{54}$ ergs.
It was a surprise to many in the astrophysics community that the
black hole masses in the first LIGO event, Eq.(\ref{LIGO}), were so large.
It was not a surprise to physicists who had studied diligently
the theory of dark matter
suggested in \cite{PF} where large numbers of black holes with many
solar masses were predicted before the LIGO announcement.

\bigskip

\noindent
Because of the no-hair theorem, there is no way of knowing whether the
initial black holes in Eq.(\ref{LIGO}) are primordial or the results of gravitational
collapse. The LIGO discovery does offer support to our theory of dark matter 
but it is premature to take this support too seriously\cite{Riess}.

\bigskip

\noindent
In order to confirm the theory of dark matter proposed in \cite{PF} the most
promising method is by an extension of the microlensing
observations reported in 2000 by the well-known MACHO Collaboration\cite{Alcock}.
That remarkable experiment was completed at the end of the twentieth
century when there was a prejudice that PBHs were mostly lighter
than the Sun. Impressive light curves were measured for microlensing
of stars in the Large Magellanic Cloud
with durations ranging from two hours to almost one year corresponding to
MACHO masses in the approximate range

\begin{equation}
10^{-5} M_{\odot} \leq M_{MACHO} \leq 25 M_{\odot}
\label{MACHO}
\end{equation} 

\noindent
What is eagerly awaited therefore are the results from a search for longer duration light curves using
the same strategy. This requires a suitable wide-angle Southern Hemisphere (to
see the LMC and SMC) telescope.
The best present choice (prior to LSST) is the Blanco 4m telescope which is fitted with a 
520-megapixel camera (DECam). Such an experiment could identify the dark matter with 
certainty.

\noindent
\section*{Acknowledgement}

\noindent
We acknowledge useful discussions with S. Sarkar and J. Silk. We thank INFN for support 
and the Physics Department at the Universitiy of Salento for hospitality.

\end{document}